\documentclass[twocolumn,showpacs,prb,amsmath,amssymb,superscriptaddress,floatfix]{revtex4} 

\usepackage{graphicx}
\usepackage{dcolumn}
\usepackage{bm}

\begin{document} 

\setlength{\topmargin}{0in}

\title{An Anomalous Phase in the Relaxor Ferroelectric 
Pb(Zn$_{1/3}$Nb$_{2/3}$)O$_3$} 
\author{Guangyong Xu}
\affiliation{Brookhaven National Laboratory, Upton, 
New York 11973}
\author{Z.~Zhong} 
\affiliation{Brookhaven National 
Laboratory, Upton, New York 11973}
\author{Y.~Bing}
\affiliation{Department of Chemistry, Simon Fraser University, Burnaby, 
British Columbia, Canada, V5A 1S6}
\author{Z.-G.~Ye}
\affiliation{Department of Chemistry, Simon Fraser University, Burnaby, 
British Columbia, Canada, V5A 1S6}
\author{C.~Stock}
\affiliation{Department of Physics, University of Toronto, Toronto, Ontario, 
Canada M5S 1A7} 
\author{G.~Shirane} 
\affiliation{Brookhaven National Laboratory, Upton, 
New York 11973}
\date{\today} 
\begin{abstract}

X-ray diffraction studies on a Pb(Zn$_{1/3}$Nb$_{2/3}$)O$_3$ (PZN)
single crystal sample show the presence of two different structures. An 
outer-layer exists in the outer most $\sim$ 10 to 50~$\mu$m of the crystal, and 
undergoes a structural phase transition at the Curie temperature
$T_C\approx410$~K. The inside
phase is however, very different. The  lattice inside the crystal 
maintains a cubic 
unit cell, while ferroelectric polarization develops 
below $T_C$. The lattice parameter of the 
cubic unit cell remains virtually a constant, i.e., much less variations  
compared to that of a typical relaxor 
ferroelectric, in a wide temperature range of 15~K to 750~K.
On the other hand, broadening of Bragg peaks and change of Bragg profile 
line-shapes in both longitudinal and transverse directions  at $T_C$ clearly 
indicate a structural phase transition occurring.

\end{abstract} 
 

\maketitle 

\section{Introduction}

\begin{figure}[ht]
\includegraphics[width=\linewidth]{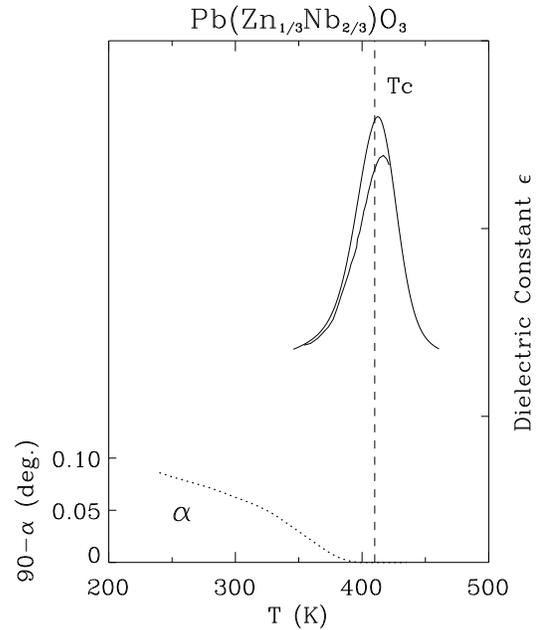}
\caption{Schematic of the frequency dependent dielectric constant $\epsilon$,
and the rhombohedral distortion angle 90$^\circ-\alpha$ (by Lebon 
{\it et al.}~\cite{Lebon}) for pure
PZN, plotted vs. temperature T. The vertical dashed line indicate the Curie 
temperature $T_C$.}
\label{fig:1}
\end{figure}

Pb(Zn$_{1/3}$Nb$_{2/3}$)O$_3$ (PZN) is a typical relaxor with a broad and 
strongly frequency-dependent dielectric constant $\epsilon$. When doped 
with PbTiO$_3$ (PT), forming solid solutions of PZN-$x$PT, the piezoelectric 
properties are greatly enhanced~\cite{PZT1,PZN_phase1,PZN_phase2}. The great 
potential for industrial applications inspired a series of studies on the PT doped 
systems. One of the most important among those was the discovery of a 
monoclinic phase which is directly related to the high piezoelectric 
response~\cite{Polarization,Universal_phase,PZN_phase,Uesu}. 
Pure PZN itself  was 
believed to undergo a cubic-to-rhombohedral phase transition upon cooling with 
zero-field at the Curie temperature $T_C\approx 410$~K. Nevertheless, 
very few direct structural studies on PZN have been reported for a long 
time. Only recently, Lebon~{\it et al.} performed the first explicit zero-field 
x-ray diffraction measurements on pure PZN single crystals~\cite{Lebon}. 
They observed the transformation into the  rhombohedral 
phase at $T_C$, as previously believed. 
As shown in Fig.~\ref{fig:1}, the phase transition occurs 
at about the same temperature where the dielectric constant $\epsilon$ peaks,
and the rhombohedral distortion develops with cooling.

However, more  recent high energy (67~keV) x-ray diffraction measurements by
Xu~{\it et al.} revealed the existence of a different phase~\cite{PZN_Xu}. 
67~keV x-ray beams can penetrate deeply
into the sample and therefore probe the inside of the crystal. 
It was found that the 
structure inside of the pure PZN single crystal has an undistorted
lattice (unit cell shape) instead. 
The true symmetry (atomic shifts) of the structure was not yet
determined, and this unidentified phase was named ``phase X''.
On the other hand, an outer-layer of about 10 to 
50~$\mu$m was found to exist and have a different structure than the inside of 
the crystal. At room temperature ($T<T_C$), the outer-layer has a rhombohedrally
distorted structure, consistent with results by Lebon~{\it et al.} This suggests
that previous x-ray diffraction measurements with x-ray energies around 10~keV
only measure the outer-layer structure, as do powder diffraction measurements,
since the average grain size of  powder samples is $\sim 50$~$\mu$m. 
Most recent x-ray studies have found similar inside/outer-layer type 
structures in PZN-4.5$\%$PT and PZN-$8\%$PT~\cite{Xu_apl}.

Phase X is extremely unusual in many aspects. 
Neutron scattering measurements by Stock {\it et al.}~\cite{Stock1} on a 
pure PZN single crystal show that the (300) Bragg 
intensity increases significantly at $T_C$ upon cooling. This can
be explained by a symmetry lowering and therefore release of extinction 
effect in the system. In addition, both longitudinal and transverse width
of the Bragg peak measured show  broadening effects at the phase transition 
temperature $T_C$. All of these indicate that although there is no rhombohedral
lattice distortion in phase X, a structural phase transition does occur
at $T_C$.

In this paper, we present detailed high energy (67~keV) x-ray diffraction 
measurements on single crystal PZN. Our results show that the phase (X) 
inside of the crystal behaves very differently than a normal ferroelectric 
phase. In spite of the absence of rhombohedral distortions, strain and 
effective sample mosaic both develop with cooling below $T_C$. However, the 
lattice 
parameter $a$ of phase X appears to be virtually a constant over a large 
temperature range of 15~K to 750~K, quite contrary to the thermal expansion of 
typical ferroelectric oxide systems.

\section{Experiment}

The PZN single crystal used in our measurements is the same unpoled 
crystal used in Ref.~\onlinecite{PZN_Xu}, 
grown at the Simon Fraiser 
University. The dimensions are $3\times3\times1$~mm$^3$. High energy 
x-ray diffraction measurements were carried out at X17B1 beamline of the 
National Synchrotron Light Source (NSLS). The x-ray energy at X17B1 
beamline is 67~keV, with an attenuation length of $\sim 400$~$\mu$m in PZN. 
The measurements were performed in the transmission mode to ensure that
the inside of the crystal is being probed. Measurements on the ``outer-layer'' 
structure were done with x-ray energy of 10.2~keV at X22A beamline of the NSLS.
In this case the x-ray attenuation length is only $\sim 10$~$\mu$m and the 
diffraction measurements were done in the reflection mode instead. 
Careful wavelength calibrations have been done on both beam lines with perfect
Si and Ge crystals. Our measurements were carried out in the temperature range 
of 15~K to 750~K. The sample holder was specially redesigned to minimize
external strain on the sample.

\section{Results}

\begin{figure}[ht]
\includegraphics[width=\linewidth]{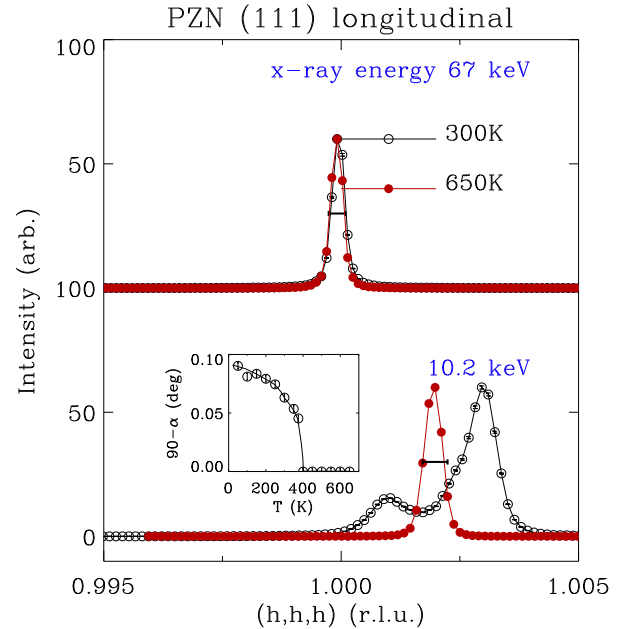}
\caption{Longitudinal scans through the (111) Bragg peak, measured at 
temperatures above and below $T_C$.  The top panel are diffraction 
results using 67~keV
x-rays, and the bottom panel with 10.2~keV x-rays. The inset shows the 
rhombohedral distortion angle derived from the 10.2~keV x-ray 
results. The horizontal bars indicate the instrument resolutions.}
\label{fig:2}
\end{figure}

In a rhombohedral phase, the lattice is stretched along the body diagonal 
direction. Therefore, the \{111\} Bragg peaks should be most sensitive to 
the rhombohedral distortion, and split due to the two \{111\} 
lengths in  different domains of the crystal. In Fig.~\ref{fig:1}, 
longitudinal scans ($\theta-2\theta$ scans) through  the (111) Bragg peak are 
shown. The units are multiples of the reciprocal lattice unit 
$a^*=2\pi/4.067$~\AA~$=1.545$~\AA$^{-1}$. When probed 
with 67~keV x-rays, the inside of the crystal shows no sign of rhombohedral 
distortion. The (111) Bragg peak remains single and sharp both above and 
below $T_C$. In contrast to neutron scattering measurements on the 
(110) and (200) Bragg peaks~\cite{Stock1}, no longitudinal broadening was 
observed by 67~keV x-ray measurements on the (111) Bragg peak. 
The three \{200\} Bragg peaks were also re-visited at room temperature 
($T<T_C$) to examine the previously observed slight tetragonal 
distortion~\cite{PZN_Xu}. The difference between the lengths of  three axis
$a$, $b$, and $c$ is less than $0.02\%$. This confirms that phase X has 
an average cubic unit cell shape ($a=b=c$), and previously observed
tetragonal distortion ($c$ slightly larger than $a$) was likely 
due to external strain created by the sample mount.

\begin{figure}[ht]
\includegraphics[width=\linewidth]{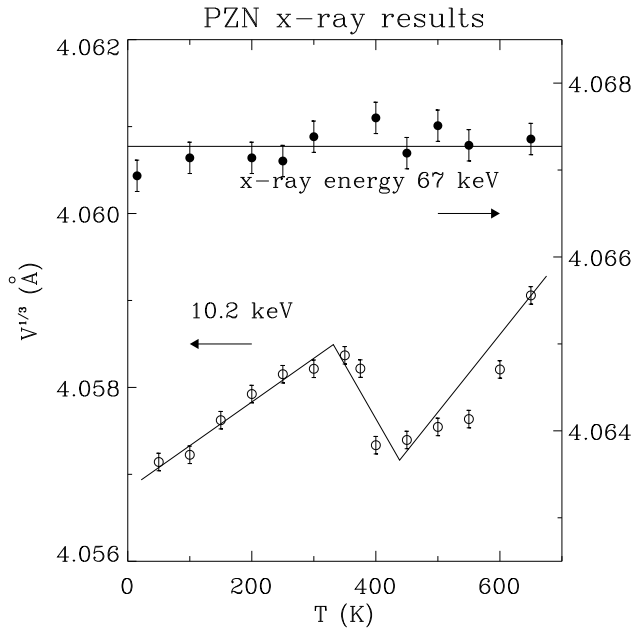}
\caption{Lattice parameter $a=Volume^{1/3}$, measured by 67~keV x-rays (inside)
and 10.2~keV x-rays (outer-layer). }
\label{fig:3}
\end{figure}

When looking at the outer-layer with 10.2~keV x-ray beams,
a splitting of the (111) peak occurs below $T_C$. Assuming a rhombohedral 
structure, the lattice parameter and rhombohedral distortion 
angle can be calculated based on the positions of the two split peaks.
As shown in the inset of Fig.~\ref{fig:2}, the rhombohedral distortion
in the outer-layer starts to appear at $T_C$, and develops upon cooling, 
consistent with previous reports from Lebon {\it et al.}~\cite{Lebon}.

\begin{figure}[ht]
\includegraphics[width=\linewidth]{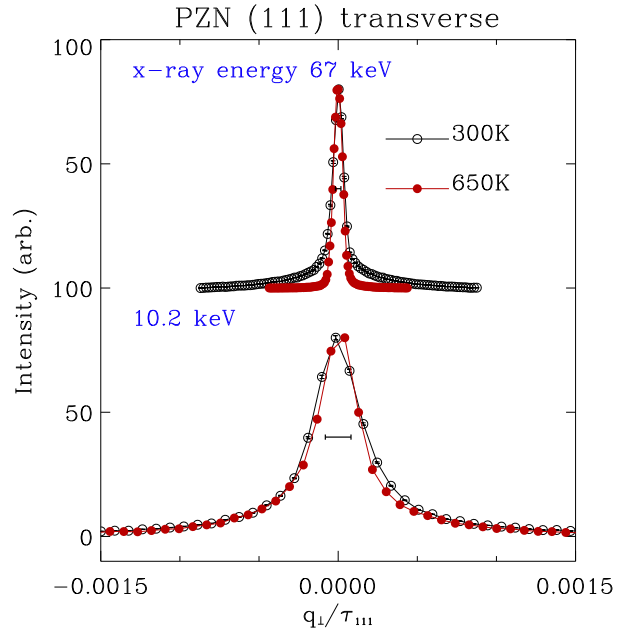}
\caption{Transverse scans through the (111) Bragg peak, measured at 
temperatures above and below $T_C$.  The top panel are diffraction 
results using 67~keV
x-rays, and the bottom panel with 10.2~keV x-rays. The horizontal bars 
indicate the instrument resolutions. }
\label{fig:4}
\end{figure}

Fig.~\ref{fig:3} shows lattice parameter $a=Volume^{1/3}$, derived from 
our measurements, assuming a cubic lattice for the inside structure,
and rhombohedral lattice for the outer-layer (below $T_C$). 
There is a $\le 0.05\%$ systematic error between 
measurements on the two beamlines due to wave-length and $2\theta_0$ 
calibrations. For normal ferroelectric oxides, the lattice parameter decreases 
almost linearly on cooling down to $T_C$, then increases in the ferroelectric 
region, and decreases again, with a smaller slope. The typical thermal 
expansion coefficient $\alpha$ is in the order of $10^{-6}$ in the high 
temperature range. This is the behavior observed
in the outer-layer structure, by our 10.2~keV x-ray measurements, as well
as Cu K$_\beta$ x-ray measurements by Lebon {\it et al.} Recent measurements
by Bing {\it et al}~\cite{Ye_PZN} have also provided more detailed information
on the structural properties of this outer-layer.
However, the temperature dependence of the lattice parameter of the inside 
structure, phase X, is quite different and very intriguing. Compared with that 
of the outer-layer, 
the lattice parameter of the inside appears to be $\sim 0.2\%$ larger. 
It also shows much less temperature variation than that of the
outer-layer. The accurate thermal expansion coefficient for the inside lattice 
cannot be determined
from our data but we can still put an upper limit that 
$|\alpha_{inside}| < 10^{-7}$ even in the high temperature range.

\begin{figure}[ht]
\includegraphics[width=\linewidth]{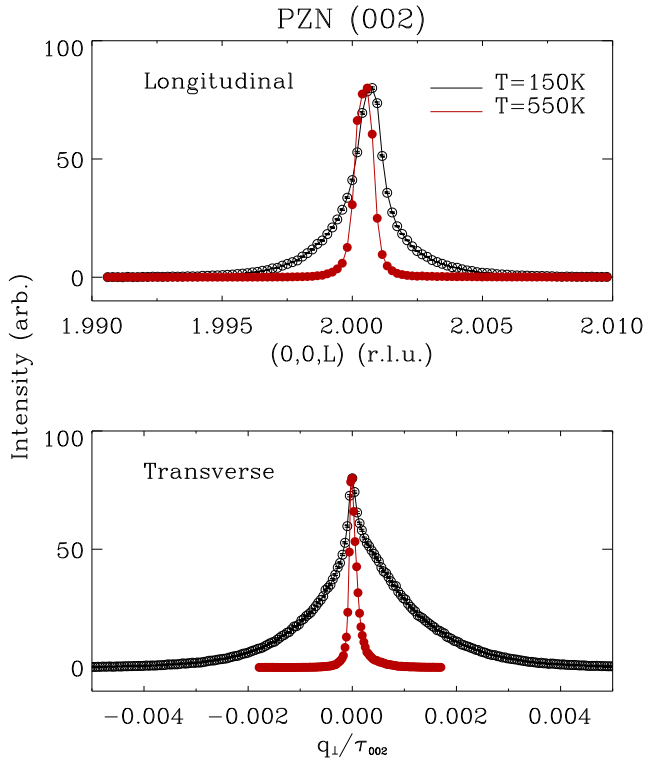}
\caption{Longitudinal and transverse scans through the (002) Bragg peak, 
measured at temperatures above and below $T_C$ with 67~keV x-rays. }
\label{fig:5}
\end{figure}

The crystal effective mosaic around the (111) Bragg peak was also 
measured by transverse scans ($\theta$ scans). Typical transverse 
scans around the (111) peak are shown in Fig.~\ref{fig:4}. With 67~keV 
x-rays probing inside of the crystal, the (111) transverse Bragg profile 
clearly changes in shape. Although the Full Width at Half Maximum (FWHM)
does not appear to change much upon cooling, the profile
apparently gains more spectral weight in the tails for T below $T_C$. In fact,
instead of the Gaussian type line shape for the high temperature cubic phase,
transverse (111) scans at T below $T_C$ in phase X yield results that can be 
better described with a Lorentzian type line shape. On the other hand, 10.2~keV 
x-ray results show very little change in the (111) transverse profiles from 
the outer-layer.

Here we note that the FWHM of the transverse scan, or, ``mosaic'' ($\eta$), 
of the outer-layer is much greater than that of the inside. At $T=300$~K, 
$\eta_{\it inside}\sim0.0048^{\circ}$, and $\eta_{\it outer layer}\sim 
0.0191^\circ$. With the instrument resolution taken out ($\delta_\theta \sim
0.002^\circ$ for 67~keV x-rays, and $\delta_\theta \sim 0.0093$ for 10.2~keV
x-rays), the effective mosaic of the outer-layer is about four times 
that of the inside. With 67~keV x-rays, although both the inside 
and the outer-layer are in the beam and contribute to the total diffraction 
intensity, the much smaller
volume (about $1\%$ to $10\%$) of the outer-layer, and more spread out of 
its diffraction intensity in $\theta$ due to the coarser mosaic, make it 
extremely difficult for the outer-layer to be detected. 
The 67~keV x-ray diffraction results are therefore dominated by those from the 
inside structure.

\begin{figure}[ht]
\includegraphics[width=\linewidth]{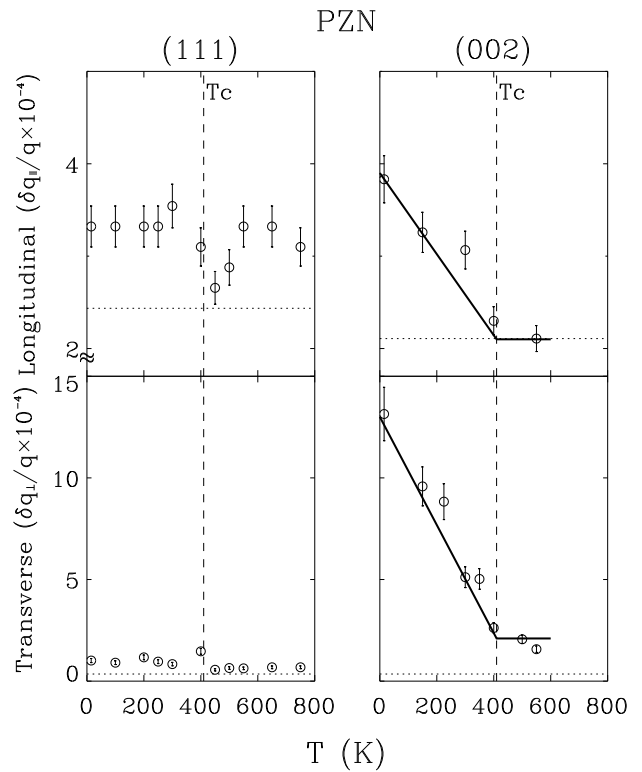}
\caption{Longitudinal and transverse Bragg widths (full width at half maximum) 
at (111) and (002) peaks, measured at different temperatures. The dotted 
horizontal lines indicate the instrument resolution, and the vertical dashed 
lines denotes the Curie temperature $T_C$.}
\label{fig:7}
\end{figure}

We have also performed similar longitudinal and transverse measurements with 
67~keV x-rays on the (002) Bragg peak of the inside structure. 
In Fig.~\ref{fig:5}, longitudinal and transverse scans through the (002) Bragg 
peak are shown, at temperatures above and below $T_C$. The temperature 
dependence of the lattice parameter agree perfectly with that obtained by
the (111) scans. Nevertheless, the change of Bragg profile width and shape
at $T_C$ becomes  more significant around the  (002) peak.  Upon cooling, in 
addition
to the broadening of the Bragg width, both longitudinal and transverse Bragg 
profile also tend to have more spectral weight 
in the tails, as observed in the (111) transverse scans.

In Fig.~\ref{fig:7}, the relative longitudinal and transverse Bragg width 
($\delta q/q$) for the (111) and (002) peaks are plotted. The Bragg width
of the (111) peak does not vary much with temperature. However, 
around the (002) peak, a clear onset of both longitudinal and transverse 
Bragg width occurs at $T_C$, and they increase monotonically upon cooling. 
The change of the Bragg profile from Gaussian shape to the Lorentzian shape
also occurs at around the same temperature. These results indicate that 
some structural change occurs at $T_C$, but not so strong as to induce 
a global lattice distortion.

\section{Discussions}

The structure of the inside of the pure PZN single crystal with an undistorted 
lattice
has also been discovered lately in another family of relaxor ferroelectrics, 
a close analog of PZN-$x$PT, PMN-$x$PT 
(Pb(Mg$_{1/3}$Nb$_{2/3}$)O$_3$-$x$PbTiO$_3$). High $q$-resolution 
neutron scattering measurements by Gehring {\it et al.} and Xu {\it et al.} 
found that phase X is also present in PMN-10$\%$PT~\cite{PMN-10PT} and 
PMN-$20\%$PT~\cite{Xu1} below the
Curie temperatures. With increasing PT concentration $x$, the low temperature
phase eventually becomes rhombohedral in PMN-27$\%$PT. These results suggest
that phase X is not an exception, but rather a universal phase which is the 
ground state for PZN-$x$PT and PMN-$x$PT systems of small PT concentrations. 
Pure PMN, which was considered to remain cubic for temperatures as low as 
5~K~\cite{Bonneau,Husson}, is likely an extreme case of phase X with 
very thin outer-layers (so that even Cu K$_\alpha$ x-rays would be able to see 
the inside undistorted structure). 

\subsection{Decoupling of polarization and lattice distortion}

Another important fact about phase X is the presence of ferroelectric 
polarization, as evidenced by the recovery of soft ferroelectric 
TO phonon below $T_C$. In the low temperature phase, the energy squared 
($(\hbar\omega)^2 \propto 1/\epsilon$) of the zone center 
TO phonon in both PMN~\cite{Waki1} and PZN~\cite{Stock1}
increases linearly upon cooling, a unique behavior of ferroelectrically
ordered system. The exact same recovery of the TO mode was also confirmed
by Raman measurements by Bovtun {\it et al.}~\cite{Raman2}. 
Note that direct measurements of the polarization - 
atomic shifts inside the crystal, are difficult because of strong extinction 
and other reasons. But many indirect measurements show results in support 
of the ferroelectric polar order below $T_C$. 
In addition to the phonon measurements, recent 
NMR~\cite{PMN_nmr} measurements also showed a clear increase of the line 
intensity corresponding to [111] Pb shifts at $T_C$, suggesting the 
development of ferroelectric polarization. 
In this phase, the internal degree of freedom of the 
structure - ferroelectric polarization, i.e., atomic shifts, is decoupled from 
the external degree of freedom - unit cell shape, that can be directly measured 
by diffraction experiments. While the polarization, as the 
primary order parameter, show signs of a phase transition into a ferroelectric
phase at $T_C$, the unit cell shape remains cubic. 
Recent reports on epitaxial films of SrTiO$_3$~\cite{SrTiO3} show similar 
decoupling effects. There the internal degree of freedom - the rotation of 
the TiO$_6$ octahedra, can be directly measured by monitoring the superlattice 
diffraction peak, and clearly shows a phase transition. Yet the unit cell 
size and shape do not show any indication of this transition. This decoupling
is possibly due to epitaxial strain and substrate clamping effects. 
The discovery of phase X is the first experimental realization of the 
decoupling between lattice distortion and ferroelectric polarization inside a 
{\it free-standing} crystal. 

One unique feature for relaxors is the existence of the polar nano-regions 
(PNR). These are nano-meter sized local polarized regions that appear  
at temperatures a few hundred degrees above $T_C$~\cite{Burns}. 
Studies on diffuse scattering intensities
indicate that the PNR are in fact displaced from the surrounding lattice. 
This was first pointed out by Hirota {\it et al.}~\cite{PMN_diffuse}, and 
the  displacement of the PNR is called the ``uniform phase shift''. This 
shift makes the PNR out-of-phase from the whole lattice, and therefore 
creates an energy barrier, that competes with the ferroelectric polarization
and prevents the whole lattice from being (rhombohedrally) distorted. More
detailed discussion about the uniform phase shift and its role on the
decoupling between the lattice distortion and ferroelectric polarization can be 
found in Ref.~\onlinecite{Xu1}. Here we would like to focus on the roles  that 
those PNR play on other properties of phase X.

\subsection{Thermal expansion}

The temperature dependence of lattice parameters for 
those systems with this undistorted phase X present are very similar to 
each other, but deviate from the typical behavior of ferroelectrics. 
The lattice parameters have been measured in pure 
PMN~\cite{PMN_Zhao}, PMN-10$\%$PT~\cite{PMN-10PT}, and PMN-20$\%$PT~\cite{Xu1}.
They remain rather constant at low temperatures,
and do not change much even above $T_C$. Only when the temperature is 
significantly 
above $T_C$, we start to see signs of clear thermal expansion. The ``onset'' 
temperature is about 500~K for PMN, 450~K for PMN-10$\%$PT, and 400~K for 
PMN-20$\%$PT, all of them are hundreds of degrees above their Curie 
temperatures ($T_C\approx 210$~K, $285$~K, and $300$~K, respectively), and 
seem to decrease with increasing PT concentration.
In our measurements on pure PZN, however, no ``onset'' has yet been observed 
(see Fig.~\ref{fig:3}). Considering 
the relatively high Curie temperature $T_C\approx410$~K, this onset temperature 
could be quite high, even beyond our current experimental capabilities (750~K). 
Taking into account the difference in $T_C$ between PZN and PMN, we find that
the temperature dependence of the lattice parameter in those two compounds are 
in close resemblance - constant lattice parameter from low temperature up to 
$\sim300$~K above $T_C$. 

The reason for this intriguing behavior is, nevertheless, not yet clear 
to us. It is in contrary to the known thermal expansion property of any 
ferroelectric oxide. On the other hand, we note that the change of lattice 
parameters of the inside structure, may not be equivalent to that of a 
macroscopic thermal expansion measurement. The outer-layer has an entirely
different temperature dependence. Another interesting 
point is that the lattice parameter of the outer-layer is actually smaller 
than that of the inside. Recent narrow beam neutron scattering 
measurements~\cite{Conlon} also indicated similar behavior (small 
outer-layer lattice parameter than the inside) in a single crystal of 
PMN-$x$PT. 
It is possible that this can create a clamping effect, 
like the substrate clamping effects on the SrTiO$_3$ thin films~\cite{SrTiO3}. 
In that case, the inside lattice would be under  external constraints, while
the whole crystal is still ``free-standing''. With the structural 
phase transition at $T_C$, the growing mismatch between the outer-layer and 
inside structures may also induce additional strain and 
stress inside the crystal.

\subsection{Structural inhomogeneity}

Phase X is also unusual in the way that structural changes occur at the 
phase transition without any macroscopic lattice distortion. Previous 
report by Mulvihill {\it et al.}~\cite{Mulvihill} suggests that below 
$T_C$, PZN could be 
in a ``microdomain'' state, with $< 0.1$~$\mu$m sized rhombohedral microdomains.
Our data indicate that those rhombohedral ``microdomains'' are very likely
to exist, but only in the ``outer-layer''. Inside the crystal, the unit
cell shape remains cubic. The Bragg peaks, especially the (002) peak, 
does broaden below $T_C$. If we attribute this broadening to the finite size
effect of ``microdomains'', the size of those domains can be estimated from
the Bragg width. For example, at $T=15$~K, the FWHM of the (002) Bragg 
peak in the longitudinal direction is $2\Gamma \sim 0.0012$~\AA$^{-1}$, 
corresponding to an average domain size of $L=0.94\pi/\Gamma\approx 5000$~\AA.
This is much larger than the size of ``microdomains''. Also consider the fact
that neutron scattering measurements~\cite{Stock1} yield much larger ($\sim 5$ 
times)
Bragg widths, after the resolution has been taken into account. It is highly 
unlikely that the ``domain size'' seen by neutrons would be much 
smaller than that seen by x-rays. 

A more feasible explanation is that the 
broadening are due to strains caused by structural inhomogeneity in the 
lattice. The source
of the  inhomogeneity is the PNR. At temperatures above $T_C$, although 
PNR already exist, the interaction between the PNR and the unpolarized 
surrounding lattice is not significant. When cooling below $T_C$, as shown 
by neutron diffuse scattering~\cite{Xu_diffuse} and specific 
heat~\cite{PMN_heat} measurements 
on PMN, both the size and total volume of the PNR increase rapidly. It is 
reasonable to believe that this is also the case in PZN. The ferroelectric 
polarization established in the whole lattice will then interact 
much more strongly with the ``out-of-phase'' PNR, trying to pull them 
back in-phase. The competition between this interaction and the energy barrier 
created by the uniform phase shift induces strain and stress in the lattice.
The lower the temperature, the stronger the interaction and the strain. 
Moreover, the PNR and this interaction is apparently not isotropic, as the 
PNR and the surrounding lattice are polarized along certain crystallographic 
orientations. This leads to the difference between strain measured at different
Bragg peaks. In our x-ray measurements, the longitudinal Bragg broadening,
which is related to the strain, is much less along the (111) direction than
the (002) direction. 
In fact, high $q$-resolution neutron scattering measurements 
have been performed on this same PZN single crystal at 300~K, on the (111) 
Bragg peak. The peak remains single, with  much smaller broadening than
that obtained by neutron measurements on (110) and (002) peaks. 

The Bragg width and lineshape also change at $T_C$ in the transverse direction.
A transverse scan is used to measure the sample ``mosaic''. Usually, the 
mosaic of a crystal is the result of defects and dislocations in the lattice,
which make the adjacent domains align at slightly different angles. 
The macroscopic result is a broadening of the crystal mosaic.
In the case of PZN, the transverse inhomogeneity, or, the effective ``mosaic'',
is more related to the PNR as 
local ``defects''. This is different from the original concept of mosaic. 
The presence of the PNR, and more importantly, their increasing
interactions with the surrounding regions, create not only strains, but also
an increasing transverse inhomogeneity with cooling. Similar to the strain, the 
transverse inhomogeneity is also anisotropic, and has strong dependence on the 
crystallographic orientations. This effective ``mosaic'' created by the PNR 
propagates through the whole crystal. In fact, if we only look at a small part 
of the crystal, only a portion of the total inhomogeneity is 
actually being measured. 
The fact that neutron measurements~\cite{Stock1} yield larger values of 
both longitudinal and transverse Bragg widths than our x-ray measurements
can be reconciled by considering the total volume of crystal in the beam.
Neutron beam sizes are much larger and usually covers the whole crystal,
and will be sensitive to the large inhomogeneity of the whole crystal.
X-ray beams, on the other hand, are much smaller (our typical beam size 
$\sim0.5$~mm in diameter), and only probes a very small part of the 
crystal. 

Overall, our high energy x-ray studies on the inside structure of single
crystal PZN reveals a phase with unique lattice properties. 
The unit cell remains cubic shape below $T_C$, while instability 
and effective mosaic in the lattice increase significantly with cooling 
at and below  the phase transition. The lattice parameter of the inside 
lattice remains almost constant for the whole temperature range from 15~K
to 750~K, in contrary to the thermal expansion behavior of any known 
ferroelectric oxide. These anomalous behaviors are not fully understood yet,
and further investigations are underway. 
There is also an outer-layer structure that is entirely 
different and behaves consistently with a typical ferroelectric system.
We speculate that the inside phase is the true  ground 
state of PZN and similar relaxor systems with low PT dopings. Its behaviors are 
much dominated by the 
interactions of the phase shifted PNR and the surrounding lattice, and 
thus highly unusual.

\begin{acknowledgments}
We would like to thank A.~A.~Bokov, P.~M.~Gehring, S.~M.~Shapiro, 
S.~B.~Vakrushev, and D. Viehland  for stimulating discussions. 
Financial support from the U.S. Department of Energy under contract 
No.~DE-AC02-98CH10886, U.S. Office of Naval Research Grant No.~N00014-99-1-0738,
and the Natural Science and Research Council of Canada 
(NSERC) is also gratefully acknowledged. 
\end{acknowledgments}


\begin{thebibliography}{27}
\expandafter\ifx\csname natexlab\endcsname\relax\def\natexlab#1{#1}\fi
\expandafter\ifx\csname bibnamefont\endcsname\relax
  \def\bibnamefont#1{#1}\fi
\expandafter\ifx\csname bibfnamefont\endcsname\relax
  \def\bibfnamefont#1{#1}\fi
\expandafter\ifx\csname citenamefont\endcsname\relax
  \def\citenamefont#1{#1}\fi
\expandafter\ifx\csname url\endcsname\relax
  \def\url#1{\texttt{#1}}\fi
\expandafter\ifx\csname urlprefix\endcsname\relax\def\urlprefix{URL }\fi
\providecommand{\bibinfo}[2]{#2}
\providecommand{\eprint}[2][]{\url{#2}}

\bibitem[{\citenamefont{Park and Shrout}(1997)}]{PZT1}
\bibinfo{author}{\bibfnamefont{S.-E.} \bibnamefont{Park}} \bibnamefont{and}
  \bibinfo{author}{\bibfnamefont{T.~R.} \bibnamefont{Shrout}},
  \bibinfo{journal}{J. Appl. Phys.} \textbf{\bibinfo{volume}{82}},
  \bibinfo{pages}{1804} (\bibinfo{year}{1997}).

\bibitem[{\citenamefont{Kuwata et~al.}(1982)\citenamefont{Kuwata, Uchino, and
  Nomura}}]{PZN_phase1}
\bibinfo{author}{\bibfnamefont{J.}~\bibnamefont{Kuwata}},
  \bibinfo{author}{\bibfnamefont{K.}~\bibnamefont{Uchino}}, \bibnamefont{and}
  \bibinfo{author}{\bibfnamefont{S.}~\bibnamefont{Nomura}},
  \bibinfo{journal}{Jpn. J. Appl. Phys.} \textbf{\bibinfo{volume}{21}},
  \bibinfo{pages}{1298} (\bibinfo{year}{1982}).

\bibitem[{\citenamefont{Kuwata et~al.}(1981)\citenamefont{Kuwata, Uchino, and
  Nomura}}]{PZN_phase2}
\bibinfo{author}{\bibfnamefont{J.}~\bibnamefont{Kuwata}},
  \bibinfo{author}{\bibfnamefont{K.}~\bibnamefont{Uchino}}, \bibnamefont{and}
  \bibinfo{author}{\bibfnamefont{S.}~\bibnamefont{Nomura}},
  \bibinfo{journal}{Ferroelectrics} \textbf{\bibinfo{volume}{37}},
  \bibinfo{pages}{579} (\bibinfo{year}{1981}).

\bibitem[{\citenamefont{Noheda et~al.}(2001)\citenamefont{Noheda, Cox, Shirane,
  Park, Cross, and Zhong}}]{Polarization}
\bibinfo{author}{\bibfnamefont{B.}~\bibnamefont{Noheda}},
  \bibinfo{author}{\bibfnamefont{D.~E.} \bibnamefont{Cox}},
  \bibinfo{author}{\bibfnamefont{G.}~\bibnamefont{Shirane}},
  \bibinfo{author}{\bibfnamefont{S.-E.} \bibnamefont{Park}},
  \bibinfo{author}{\bibfnamefont{L.~E.} \bibnamefont{Cross}}, \bibnamefont{and}
  \bibinfo{author}{\bibfnamefont{Z.}~\bibnamefont{Zhong}},
  \bibinfo{journal}{Phy. Rev. Lett.} \textbf{\bibinfo{volume}{86}},
  \bibinfo{pages}{3891} (\bibinfo{year}{2001}).

\bibitem[{\citenamefont{Cox et~al.}(2001)\citenamefont{Cox, Noheda, Shirane,
  Uesu, Fujishiro, and Yamada}}]{Universal_phase}
\bibinfo{author}{\bibfnamefont{D.~E.} \bibnamefont{Cox}},
  \bibinfo{author}{\bibfnamefont{B.}~\bibnamefont{Noheda}},
  \bibinfo{author}{\bibfnamefont{G.}~\bibnamefont{Shirane}},
  \bibinfo{author}{\bibfnamefont{Y.}~\bibnamefont{Uesu}},
  \bibinfo{author}{\bibfnamefont{K.}~\bibnamefont{Fujishiro}},
  \bibnamefont{and} \bibinfo{author}{\bibfnamefont{Y.}~\bibnamefont{Yamada}},
  \bibinfo{journal}{Appl. Phys. Lett.} \textbf{\bibinfo{volume}{79}},
  \bibinfo{pages}{400} (\bibinfo{year}{2001}).

\bibitem[{\citenamefont{La-Orauttapong
  et~al.}(2002)\citenamefont{La-Orauttapong, Noheda, Ye, Gehring, Toulouse,
  Cox, and Shirane}}]{PZN_phase}
\bibinfo{author}{\bibfnamefont{D.}~\bibnamefont{La-Orauttapong}},
  \bibinfo{author}{\bibfnamefont{B.}~\bibnamefont{Noheda}},
  \bibinfo{author}{\bibfnamefont{Z.-G.} \bibnamefont{Ye}},
  \bibinfo{author}{\bibfnamefont{P.~M.} \bibnamefont{Gehring}},
  \bibinfo{author}{\bibfnamefont{J.}~\bibnamefont{Toulouse}},
  \bibinfo{author}{\bibfnamefont{D.~E.} \bibnamefont{Cox}}, \bibnamefont{and}
  \bibinfo{author}{\bibfnamefont{G.}~\bibnamefont{Shirane}},
  \bibinfo{journal}{Phys. Rev. B} \textbf{\bibinfo{volume}{65}},
  \bibinfo{pages}{144101} (\bibinfo{year}{2002}).

\bibitem[{\citenamefont{Uesu et~al.}(2002)\citenamefont{Uesu, Matsuda, Yamada,
  Fujishiro, Cox, Noheda, and Shirane}}]{Uesu}
\bibinfo{author}{\bibfnamefont{Y.}~\bibnamefont{Uesu}},
  \bibinfo{author}{\bibfnamefont{M.}~\bibnamefont{Matsuda}},
  \bibinfo{author}{\bibfnamefont{Y.}~\bibnamefont{Yamada}},
  \bibinfo{author}{\bibfnamefont{K.}~\bibnamefont{Fujishiro}},
  \bibinfo{author}{\bibfnamefont{D.~E.} \bibnamefont{Cox}},
  \bibinfo{author}{\bibfnamefont{B.}~\bibnamefont{Noheda}}, \bibnamefont{and}
  \bibinfo{author}{\bibfnamefont{G.}~\bibnamefont{Shirane}},
  \bibinfo{journal}{J. Phys. Soc. Jpn.} \textbf{\bibinfo{volume}{71}},
  \bibinfo{pages}{960} (\bibinfo{year}{2002}).

\bibitem[{\citenamefont{Lebon et~al.}(2002)\citenamefont{Lebon, Dammak,
  Calvarin, and Ahmedou}}]{Lebon}
\bibinfo{author}{\bibfnamefont{A.}~\bibnamefont{Lebon}},
  \bibinfo{author}{\bibfnamefont{H.}~\bibnamefont{Dammak}},
  \bibinfo{author}{\bibfnamefont{G.}~\bibnamefont{Calvarin}}, \bibnamefont{and}
  \bibinfo{author}{\bibfnamefont{I.~O.} \bibnamefont{Ahmedou}},
  \bibinfo{journal}{J. Phys.: Condens. Matter} \textbf{\bibinfo{volume}{14}},
  \bibinfo{pages}{7035} (\bibinfo{year}{2002}).

\bibitem[{\citenamefont{Xu et~al.}(2003{\natexlab{a}})\citenamefont{Xu, Zhong,
  Bing, Ye, Stock, and Shirane}}]{PZN_Xu}
\bibinfo{author}{\bibfnamefont{G.}~\bibnamefont{Xu}},
  \bibinfo{author}{\bibfnamefont{Z.}~\bibnamefont{Zhong}},
  \bibinfo{author}{\bibfnamefont{Y.}~\bibnamefont{Bing}},
  \bibinfo{author}{\bibfnamefont{Z.-G.} \bibnamefont{Ye}},
  \bibinfo{author}{\bibfnamefont{C.}~\bibnamefont{Stock}}, \bibnamefont{and}
  \bibinfo{author}{\bibfnamefont{G.}~\bibnamefont{Shirane}},
  \bibinfo{journal}{Phys. Rev. B} \textbf{\bibinfo{volume}{67}},
  \bibinfo{pages}{104102} (\bibinfo{year}{2003}{\natexlab{a}}).

\bibitem[{\citenamefont{Xu et~al.}(2004{\natexlab{a}})\citenamefont{Xu, Hiraka,
  Ohwada, and Shirane}}]{Xu_apl}
\bibinfo{author}{\bibfnamefont{G.}~\bibnamefont{Xu}},
  \bibinfo{author}{\bibfnamefont{H.}~\bibnamefont{Hiraka}},
  \bibinfo{author}{\bibfnamefont{K.}~\bibnamefont{Ohwada}}, \bibnamefont{and}
  \bibinfo{author}{\bibfnamefont{G.}~\bibnamefont{Shirane}}
  (\bibinfo{year}{2004}{\natexlab{a}}), \eprint{cond-mat/0401437}.

\bibitem[{\citenamefont{Stock et~al.}(2004)\citenamefont{Stock, Birgeneau,
  Wakimoto, Gardner, Chen, Ye, and Shirane}}]{Stock1}
\bibinfo{author}{\bibfnamefont{C.}~\bibnamefont{Stock}},
  \bibinfo{author}{\bibfnamefont{R.~J.} \bibnamefont{Birgeneau}},
  \bibinfo{author}{\bibfnamefont{S.}~\bibnamefont{Wakimoto}},
  \bibinfo{author}{\bibfnamefont{J.~S.} \bibnamefont{Gardner}},
  \bibinfo{author}{\bibfnamefont{W.}~\bibnamefont{Chen}},
  \bibinfo{author}{\bibfnamefont{Z.-G.} \bibnamefont{Ye}}, \bibnamefont{and}
  \bibinfo{author}{\bibfnamefont{G.}~\bibnamefont{Shirane}},
  \bibinfo{journal}{Phys. Rev. B} \textbf{\bibinfo{volume}{69}},
  \bibinfo{pages}{094104} (\bibinfo{year}{2004}).

\bibitem[{\citenamefont{Bing et~al.}(2004)\citenamefont{Bing, Bokov, Ye,
  Noheda, and Shirane}}]{Ye_PZN}
\bibinfo{author}{\bibfnamefont{Y.-H.} \bibnamefont{Bing}},
  \bibinfo{author}{\bibfnamefont{A.~A.} \bibnamefont{Bokov}},
  \bibinfo{author}{\bibfnamefont{Z.-G.} \bibnamefont{Ye}},
  \bibinfo{author}{\bibfnamefont{B.}~\bibnamefont{Noheda}}, \bibnamefont{and}
  \bibinfo{author}{\bibfnamefont{G.}~\bibnamefont{Shirane}}
  (\bibinfo{year}{2004}), \bibinfo{note}{unpublished}.

\bibitem[{\citenamefont{Gehring et~al.}(2003)\citenamefont{Gehring, Chen, Ye,
  and Shirane}}]{PMN-10PT}
\bibinfo{author}{\bibfnamefont{P.~M.} \bibnamefont{Gehring}},
  \bibinfo{author}{\bibfnamefont{W.}~\bibnamefont{Chen}},
  \bibinfo{author}{\bibfnamefont{Z.-G.} \bibnamefont{Ye}}, \bibnamefont{and}
  \bibinfo{author}{\bibfnamefont{G.}~\bibnamefont{Shirane}}
  (\bibinfo{year}{2003}), \eprint{cond-mat/0304289}.

\bibitem[{\citenamefont{Xu et~al.}(2003{\natexlab{b}})\citenamefont{Xu,
  Viehland, Li, Gehring, and Shirane}}]{Xu1}
\bibinfo{author}{\bibfnamefont{G.}~\bibnamefont{Xu}},
  \bibinfo{author}{\bibfnamefont{D.}~\bibnamefont{Viehland}},
  \bibinfo{author}{\bibfnamefont{J.~F.} \bibnamefont{Li}},
  \bibinfo{author}{\bibfnamefont{P.~M.} \bibnamefont{Gehring}},
  \bibnamefont{and} \bibinfo{author}{\bibfnamefont{G.}~\bibnamefont{Shirane}},
  \bibinfo{journal}{Phys. Rev. B} \textbf{\bibinfo{volume}{68}},
  \bibinfo{pages}{212410} (\bibinfo{year}{2003}{\natexlab{b}}).

\bibitem[{\citenamefont{Bonneau et~al.}(1989)\citenamefont{Bonneau, Garnier,
  Husson, and Morell}}]{Bonneau}
\bibinfo{author}{\bibfnamefont{P.}~\bibnamefont{Bonneau}},
  \bibinfo{author}{\bibfnamefont{P.}~\bibnamefont{Garnier}},
  \bibinfo{author}{\bibfnamefont{E.}~\bibnamefont{Husson}}, \bibnamefont{and}
  \bibinfo{author}{\bibfnamefont{A.}~\bibnamefont{Morell}},
  \bibinfo{journal}{Mater. Re. Bull} \textbf{\bibinfo{volume}{24}},
  \bibinfo{pages}{201} (\bibinfo{year}{1989}).

\bibitem[{\citenamefont{{de\ Mathan} et~al.}(1991)\citenamefont{{de\ Mathan},
  Husson, Calvarin, Gavarri, Hewat, and Morell}}]{Husson}
\bibinfo{author}{\bibfnamefont{N.}~\bibnamefont{{de\ Mathan}}},
  \bibinfo{author}{\bibfnamefont{E.}~\bibnamefont{Husson}},
  \bibinfo{author}{\bibfnamefont{G.}~\bibnamefont{Calvarin}},
  \bibinfo{author}{\bibfnamefont{J.~R.} \bibnamefont{Gavarri}},
  \bibinfo{author}{\bibfnamefont{A.~W.} \bibnamefont{Hewat}}, \bibnamefont{and}
  \bibinfo{author}{\bibfnamefont{A.}~\bibnamefont{Morell}},
  \bibinfo{journal}{J. Phys. Condens. Matter} \textbf{\bibinfo{volume}{3}},
  \bibinfo{pages}{8159} (\bibinfo{year}{1991}).

\bibitem[{\citenamefont{Wakimoto et~al.}(2002)\citenamefont{Wakimoto, Stock,
  Birgeneau, Ye, Chen, Buyers, Gehring, and Shirane}}]{Waki1}
\bibinfo{author}{\bibfnamefont{S.}~\bibnamefont{Wakimoto}},
  \bibinfo{author}{\bibfnamefont{C.}~\bibnamefont{Stock}},
  \bibinfo{author}{\bibfnamefont{R.~J.} \bibnamefont{Birgeneau}},
  \bibinfo{author}{\bibfnamefont{Z.-G.} \bibnamefont{Ye}},
  \bibinfo{author}{\bibfnamefont{W.}~\bibnamefont{Chen}},
  \bibinfo{author}{\bibfnamefont{W.~J.~L.} \bibnamefont{Buyers}},
  \bibinfo{author}{\bibfnamefont{P.~M.} \bibnamefont{Gehring}},
  \bibnamefont{and} \bibinfo{author}{\bibfnamefont{G.}~\bibnamefont{Shirane}},
  \bibinfo{journal}{Phys. Rev. B} \textbf{\bibinfo{volume}{65}},
  \bibinfo{pages}{172105} (\bibinfo{year}{2002}).

\bibitem[{\citenamefont{Bovtun et~al.}(2003)\citenamefont{Bovtun, Kamba,
  Pashkin, and Savinov}}]{Raman2}
\bibinfo{author}{\bibfnamefont{V.}~\bibnamefont{Bovtun}},
  \bibinfo{author}{\bibfnamefont{S.}~\bibnamefont{Kamba}},
  \bibinfo{author}{\bibfnamefont{A.}~\bibnamefont{Pashkin}}, \bibnamefont{and}
  \bibinfo{author}{\bibfnamefont{M.}~\bibnamefont{Savinov}}
  (\bibinfo{year}{2003}), \bibinfo{note}{proceedings of NATO Advanced Research
  Workshop on the Disordered Ferroelectrics, Kiev}.

\bibitem[{\citenamefont{Blinc et~al.}(2003)\citenamefont{Blinc, Laguta, and
  Zalar}}]{PMN_nmr}
\bibinfo{author}{\bibfnamefont{R.}~\bibnamefont{Blinc}},
  \bibinfo{author}{\bibfnamefont{V.}~\bibnamefont{Laguta}}, \bibnamefont{and}
  \bibinfo{author}{\bibfnamefont{B.}~\bibnamefont{Zalar}},
  \bibinfo{journal}{Phys. Rev. Lett.} \textbf{\bibinfo{volume}{91}},
  \bibinfo{pages}{247601} (\bibinfo{year}{2003}).

\bibitem[{\citenamefont{He et~al.}(2003)\citenamefont{He, Wells, Shapiro, .v.
  Zimmermann, Clark, and Xi}}]{SrTiO3}
\bibinfo{author}{\bibfnamefont{F.}~\bibnamefont{He}},
  \bibinfo{author}{\bibfnamefont{B.~O.} \bibnamefont{Wells}},
  \bibinfo{author}{\bibfnamefont{S.~M.} \bibnamefont{Shapiro}},
  \bibinfo{author}{\bibfnamefont{M.}~\bibnamefont{.v. Zimmermann}},
  \bibinfo{author}{\bibfnamefont{A.}~\bibnamefont{Clark}}, \bibnamefont{and}
  \bibinfo{author}{\bibfnamefont{X.~X.} \bibnamefont{Xi}},
  \bibinfo{journal}{Appl. Phys. Lett.} \textbf{\bibinfo{volume}{83}},
  \bibinfo{pages}{123} (\bibinfo{year}{2003}).

\bibitem[{\citenamefont{Burns and Dacol}(1983)}]{Burns}
\bibinfo{author}{\bibfnamefont{G.}~\bibnamefont{Burns}} \bibnamefont{and}
  \bibinfo{author}{\bibfnamefont{F.~H.} \bibnamefont{Dacol}},
  \bibinfo{journal}{Phys. Rev. B} \textbf{\bibinfo{volume}{28}},
  \bibinfo{pages}{2527} (\bibinfo{year}{1983}).

\bibitem[{\citenamefont{Hirota et~al.}(2002)\citenamefont{Hirota, Ye, Wakimoto,
  Gehring, and Shirane}}]{PMN_diffuse}
\bibinfo{author}{\bibfnamefont{K.}~\bibnamefont{Hirota}},
  \bibinfo{author}{\bibfnamefont{Z.-G.} \bibnamefont{Ye}},
  \bibinfo{author}{\bibfnamefont{S.}~\bibnamefont{Wakimoto}},
  \bibinfo{author}{\bibfnamefont{P.~M.} \bibnamefont{Gehring}},
  \bibnamefont{and} \bibinfo{author}{\bibfnamefont{G.}~\bibnamefont{Shirane}},
  \bibinfo{journal}{Phys. Rev. B} \textbf{\bibinfo{volume}{65}},
  \bibinfo{pages}{104105} (\bibinfo{year}{2002}).

\bibitem[{\citenamefont{Zhao et~al.}(1998)\citenamefont{Zhao, Glazounov, Zhang,
  and Toby}}]{PMN_Zhao}
\bibinfo{author}{\bibfnamefont{J.}~\bibnamefont{Zhao}},
  \bibinfo{author}{\bibfnamefont{A.~E.} \bibnamefont{Glazounov}},
  \bibinfo{author}{\bibfnamefont{Q.~M.} \bibnamefont{Zhang}}, \bibnamefont{and}
  \bibinfo{author}{\bibfnamefont{B.}~\bibnamefont{Toby}},
  \bibinfo{journal}{Appl. Phys. Lett.} \textbf{\bibinfo{volume}{72}},
  \bibinfo{pages}{1048} (\bibinfo{year}{1998}).

\bibitem[{\citenamefont{Conlon and Stock}()}]{Conlon}
\bibinfo{author}{\bibfnamefont{K.}~\bibnamefont{Conlon}} \bibnamefont{and}
  \bibinfo{author}{\bibfnamefont{C.}~\bibnamefont{Stock}},
  \bibinfo{note}{private communications}.

\bibitem[{\citenamefont{Mulvihill et~al.}(1997)\citenamefont{Mulvihill, Cross,
  Cao, and Uchino}}]{Mulvihill}
\bibinfo{author}{\bibfnamefont{M.~L.} \bibnamefont{Mulvihill}},
  \bibinfo{author}{\bibfnamefont{L.~E.} \bibnamefont{Cross}},
  \bibinfo{author}{\bibfnamefont{W.}~\bibnamefont{Cao}}, \bibnamefont{and}
  \bibinfo{author}{\bibfnamefont{K.}~\bibnamefont{Uchino}},
  \bibinfo{journal}{J. Am. Ceram. Soc} \textbf{\bibinfo{volume}{80}},
  \bibinfo{pages}{1462} (\bibinfo{year}{1997}).

\bibitem[{\citenamefont{Xu et~al.}(2004{\natexlab{b}})\citenamefont{Xu,
  Shirane, Copley, and Gehring}}]{Xu_diffuse}
\bibinfo{author}{\bibfnamefont{G.}~\bibnamefont{Xu}},
  \bibinfo{author}{\bibfnamefont{G.}~\bibnamefont{Shirane}},
  \bibinfo{author}{\bibfnamefont{J.~R.~D.} \bibnamefont{Copley}},
  \bibnamefont{and} \bibinfo{author}{\bibfnamefont{P.~M.}
  \bibnamefont{Gehring}}, \bibinfo{journal}{Phys. Rev. B}
  \textbf{\bibinfo{volume}{69}}, \bibinfo{pages}{064112}
  (\bibinfo{year}{2004}{\natexlab{b}}).

\bibitem[{\citenamefont{Moriya et~al.}(2003)\citenamefont{Moriya, Kawaji, Tojo,
  and Atake}}]{PMN_heat}
\bibinfo{author}{\bibfnamefont{Y.}~\bibnamefont{Moriya}},
  \bibinfo{author}{\bibfnamefont{H.}~\bibnamefont{Kawaji}},
  \bibinfo{author}{\bibfnamefont{T.}~\bibnamefont{Tojo}}, \bibnamefont{and}
  \bibinfo{author}{\bibfnamefont{T.}~\bibnamefont{Atake}},
  \bibinfo{journal}{Phys. Rev. Lett.} \textbf{\bibinfo{volume}{90}},
  \bibinfo{pages}{205901} (\bibinfo{year}{2003}).

\end{thebibliography}

\end{document}